# Classical polarimetry with a twist:
## *a compact, geometric approach*

**William B. Sparks,[1,*] Thomas A. Germer[2] , Rebecca M. Sparks[3]**

[1]*SETI Institute, 189 Bernardo Avenue, Suite 200, Mountain View, CA 94043*
[2]*National Institute of Standards and Technology, 100 Bureau Drive, Gaithersburg, MD 20899*
[3]*Irvine Nature Center, 11201 Garrison Forest Road, Owings Mills, MD 21117*

[*]*Corresponding author: wsparks@seti.org*

**Abstract:** We present an approach to classical polarimetry that requires no moving parts, is compact and robust, and that encodes the complete polarization information on a single data frame, accomplished by replacing the rotation of components such as wave plates with position along a spatial axis. We demonstrate the concept with a polarimeter having a quarter wave plate whose fast axis direction changes with location along one axis of a 2D data frame in conjunction with a fixed-direction polarization analyzer, analogous to a classical rotating quarter wave plate polarimeter. The full set of Stokes parameters is obtained, with maximal sensitivity to circular polarization Stokes *V* if a quarter wave retarder is used. Linear and circular polarization terms are encoded with spatial carrier frequencies that differ by a factor two, which minimizes cross-talk. Other rotating component polarimeters lend themselves to the approach. Since the polarization modulation spatial frequencies do not change greatly, if at all, with wavelength such devices are close to achromatic, simplifying instrument design. Since the polarimetric information is acquired in a single observation, rapidly varying, transient and moving targets are accessible, loss of precision due to sequential data acquisition is avoided, and moving parts are not required.

## 1.  Introduction

Measurement of the polarization of light provides a diverse suite of diagnostics useful across the physical and biological sciences. From medical applications to the non-thermal emission of distant quasars, polarimetry offers insights and capabilities unavailable to spectroscopy. An application of interest to the authors is the use of circular polarization as a remote sensing diagnostic of homochirality, and hence of the presence of microbial life beyond the Earth (Sparks et al. 2009; Sparks et al. 2015; Patty et al. 2018). In astronomy and the Earth sciences, polarimetry is typically used to study interstellar dust, particle aerosols, clouds and solid surfaces. The National Aeronautics and Space Administration (NASA) has, for example, directed the Plankton, Aerosol, Cloud, ocean Ecosystem Earth sciences space mission[1] (PACE) to include at least one polarimeter. Measurement of polarization can be hampered by the properties of the target, which may be moving, varying rapidly, be too bright or too faint, be hidden in the radiation of a contaminating source, or it may be hampered by the properties of the detection system, which can be large, bulky, expensive, fragile, often requiring sequential acquisition of data.

The concept of obtaining polarization information encoded onto either spatial or spectral data dimensions is often referred to as channeled or single-shot polarimetry. A variety of such methods is reviewed by Goldstein (2011). Snik et al. (2009) describe an instrument for spectral channeled polarimetry (SPEX) in which the polarization information is encoded as an intensity modulation along the spectrum of the target, as a function of wavelength, and Oka et al. (2013) consider a mixture of approaches to channeled polarimetry. Snik et al. (2014) generalize the multitude of approaches using a multidomain modulation conceptualization. Sparks et al. (2012) showed that full Stokes spectropolarimetry is possible using a single two-dimensional data frame, with classical spectroscopy along one dimension, and the polarization encoded as an intensity modulation along the orthogonal dimension (hereafter, "polarimetric dimension") on the detector. That method uses a spatially variable retardance obtained by inserting two opposing birefringent wedges with orthogonal fast axes. Two such wedge pairs with different pitch angles encode the full Stokes vector (Sparks et al. 2012; Pertenais et al. 2015a; b)[2].

---
[1] https://pace.gsfc.nasa.gov/, https://pace.oceansciences.org/mission.htm

*Here we explore a different approach: instead of using a variable retardance along the polarimetric dimension, we use a (spatially) varying retarder fast axis direction*[3]. The resulting data is similar to that of the birefringent wedges concept, in that intensity modulation orthogonal to the spectral dimension encodes the Stokes parameters as coefficients of orthogonal trigonometric functions; however, it differs in a number of important ways. One of the differences is its achromaticity, and another is the simplicity of the trigonometric functions. The spatially variable retardance and variable fast axis (or geometric) polarimetry concepts are physically analogous to two approaches available for classical precision polarimeters: modulate the retardance using dual photoelastic modulators (PEMs) or nematic liquid crystals (e.g., Hough et al. 2006) or modulate the retarder fast axis direction using ferroelectric liquid crystals (e.g., Bailey et al. 2015), though in both cases now without the need to modulate in time. The geometric approach can be generalized to a variety of polarimeter implementations based upon rotating polarization elements, with the component rotation replaced by a static mapping along a spatial or polarimetric dimension, and, like the dual PEM, in the quarter wave implementation the carrier frequencies for circular and linear Stokes parameters differ by a factor two.

An example of a concept using space-variant polarization state manipulation is the half wave plate vortex retarder, available off-the-shelf commercially and implemented in astronomy as an essential element in certain high contrast imaging coronagraph designs (Mawet et al. 2017; Serabyn et al. 2017). The phase delay of the retarder is constant across the optical element, half wave in this example, but the direction of the fast axis rotates with azimuth around a point. The "charge" $m$ on the vortex retarder refers to the number of complete rotations of the fast axis as the azimuth changes through $2\pi$ radians. A variety of technologies are available to construct such devices, for example photoaligned liquid crystal polymers provide a versatile option (McEldowney et al. 2008). Below, we illustrate the geometric concept in polarimetry with a simple, macroscopic analog constructed from inexpensive, discrete components.

By using a quarter wave plate with a changing fast axis direction along the polarimetric dimension of the 2D data, the polarimeter is essentially identical to a full Stokes rotating quarter wave polarimeter (e.g., Azzam 1978; Goldstein 1992, 2011; Arnoldt 2011; Appendix A), though neither sequential measurements nor rotating elements are required. For the specific example of detecting circular polarization as evidence of homochirality, the detection sensitivity to circular polarization is higher than it is to linear polarization, and the information is carried on a spatial frequency differing by a factor two, hence cross talk is minimal or non-existent (Appendix A).

With a half wave geometrically variable retarder, followed by a polarization analyzer, the polarimeter encodes just linear polarization. With polarization analyzers whose direction changes along the polarimetric dimension instead of rotating, a variety of alternative configurations also emerge (Goldstein 2011), see Appendix B.

---

[2] Patents: US 9097585 B2; PCT/US2016/028380
[3] Patent pending

The advantages of a polarimeter which provides intensity modulation in this way on a single data frame are: it is small, compact, robust, requires no moving parts, can encode the full set of Stokes parameters, has high dynamic range, and is achromatic. Polarimetry obtained on a single data frame may be applied to the normal suite of polarimetric targets, but in addition, moving, time varying and transient sources are accessible. The polarimetric (spatial), modulation is independent of the spectroscopy hence the intrinsic spectral resolution of the spectrometer is preserved on demodulation of the polarization. Either high or low spectral resolutions can be used. By spreading the light across the detector along the polarimetric dimension, there are significant benefits to the flat-fielding. If residual flat-field errors remain, the influence of individual pixel blemishes is diminished by the low-order demodulation, similar to the benefits of averaging over many pixels in photometry, while a dual beam implementation guards against additional systematic concerns, Appendix A.

Given their potential capacity to enable detection of extrasolar planets orbiting nearby stars, efforts are under way to space qualify vortex coronagraphs, while Solar space missions have space qualified electro-optical components for polarimetry (Capobianco et al. 2012; Fineschi et al. 2012). Hence space application is feasible for this class of polarimeters exploiting similar technologies, with the advantages relative to conventional polarimeters of no moving or electronically modulating parts, compactness, and simplicity of design and analysis, as well as access to time-variable science.

## 2. Theoretical framework

There are numerous ways to measure the polarization of light (e.g., Goldstein 2011). One of the most classical and widely used methods is a rotating retarder, or modulator, with a polarizer. We assume a quarter wave retarder here. Polarization measurement can be accomplished with a variety of discrete configurations, including removal of the quarter-wave plate, or by rotation, continuous or otherwise, of the quarter-wave plate, or rotation of the polarizer. Goldstein (2011) describes these and other techniques for accomplishing Stokes polarimetry, many of which depend on sequential data acquisition. Conventional techniques to eliminate the time dependence typically require division of the incoming beam into multiple beams and multiple detectors, with consequent difficulties of spatial or temporal scene matching, or the encoding and modulation methods of channeled polarimetry.

Here, to illustrate the concept of a geometric approach, we focus on the quarter wave polarimeter which, like its classical counterpart, has the promise to be a powerful and versatile polarimeter in its own right. There is an extensive literature discussing numerous aspects of the quarter wave plate polarimeter and its optimization, and optimization of polarimeter performance in general (del Torro Iniesta & Collados 2000). In a rotating quarter wave plate Stokes polarimeter, an incoming light beam encounters a quarter-wave plate prior to a fixed linear polarizer and detector. The quarter-wave fast axis direction can be set to different angles, or rotated, with respect to the transmission axis of the linear polarizer. By using a spatially varying fast axis direction along the slit of a conventional long slit spectrograph, together with a fixed linear polarization analyzer (which can be a single polarizer or polarizing beam splitter to preserve flux, provide redundancy and eliminate certain systematic errors), we achieve an intensity modulation orthogonal to the spectrum, along the polarimetric dimension, which encodes the full Stokes vector as coefficients of simple, orthogonal trigonometric functions, Fig. 1. The maximum sensitivity is to circular polarization. If a similarly configured half-wave plate is used instead of a quarter-wave plate, the modulation encodes the linear polarization Stokes parameters only, while a general retardance permits optimization or balancing of the sensitivities for the different Stokes parameters (Sabatke et al. 2000).

Let an incoming light beam be described by its Stokes vector $S \equiv (I, Q, U, V) \equiv I(1, q, u, v)$ where I is the total intensity, $Q, U, V$ are the unnormalized Stokes parameters, and $q, u, v$ are the normalized Stokes parameters. The degree of polarization is $p = \sqrt{q^2 + u^2 + v^2}$, and the linear polarization position angle is $\zeta = \frac{1}{2}\tan^{-1}(u/q)$. The intensity modulation of the output as a function of retarder fast axis angle $\theta$ (equated to position along the polarimetric axis orthogonal to the spectrum, either discrete as a set of $n$ angles $\theta_i$ or continuous) for an ideal quarter-wave plate plus ideal single element polarizer is given by (Goldstein 2011, equation 14.64; Appendix A):

$$I_{\text{obs}}(\theta_i) = \frac{1}{2n}\left(I + \frac{1}{2}Q + \frac{1}{2}Q\cos 4\theta_i + \frac{1}{2}U\sin 4\theta_i - V\sin 2\theta_i\right).$$

For an ideal half wave plate plus ideal single element polarizer:

$$I_{\text{obs}}(\theta_i) = \frac{1}{2n}(I + Q\cos 4\theta_i + U\sin 4\theta_i).$$

That is, the circular polarization Stokes $V$ is given by the coefficient of sin $2\theta$, while $Q$ and $U$ are given by the coefficients of sin $4\theta$ and cos $4\theta$. For convenience, a Mueller matrix mathematical analysis is presented in Appendix A. The relationships are robust against departures from exact quarter wave retardance, and inefficiencies in the polarization analyzer, as shown in Appendix A.

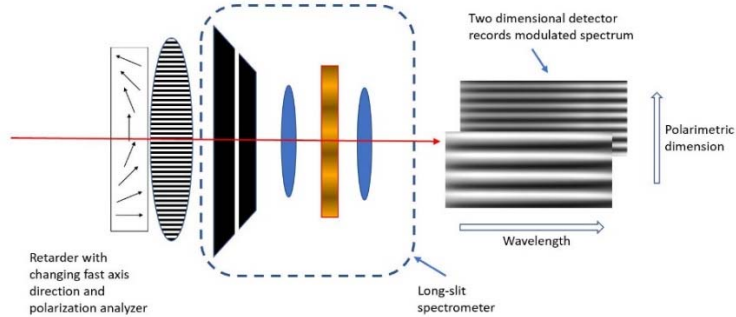

Fig. 1. Concept illustration: incoming beam from the left encounters a quarter wave plate with spatially varying fast axis direction, followed by a polarization analyzer and long-slit spectrometer. The 2D detector frame illustrates circular or linear polarization, linear having twice the modulation frequency.

## 3. Practical implementation

A polarization component with a spatially varying fast axis can be manufactured using a variety of methods. With microscale formulation of the polarizing element utilizing ultraviolet cured polymers and/or liquid crystal techniques, a retarder may be generated in which, to all intents and purposes, the fast axis of the wave plate changes continuously with position along one axis of the element (McEldowney et al. 2008; Komanduri et al. 2013; Miskiewicz & Escuti 2014). In the case of the vector vortex coronagraph, for example, the wave plate has a constant retardance, but the fast axis direction changes with azimuth around a central point or singularity. With similar manufacturing methods, it is possible to generate a smoothly varying fast axis position angle with position along a cartesian axis, $x$ or $y$, and with configurable levels of achromaticity with wavelength (Komanduri et al. 2013).

Here, we illustrate the concept by assembling a simple optical element using a discrete set of quarter-wave plate strips cut from inexpensive polymer at a set of appropriate fast-axis angles, reoriented so the strips are parallel. The device functions as a full Stokes polarimeter, and yields surprisingly high-quality results, although the intent is simply to show polarization sensitivity and functional behavior, as a proof of concept. Hence, we did not adopt any calibration procedures beyond the most basic: the data frames were debiased and flat-fielded and spectra were extracted following identification of regions unaffected by scattered light, and normalized to the white-light spectrum. Demodulation equations as given in Appendix A were used, with a small correction to the nominal strip-angles.

Eight narrow strips somewhat less than 1 mm in width and a few cm in length were cut by hand from quarter-wave retarder film at position angles corresponding to 22.5° increments, from the horizontal, to 157.5°. The strips were reoriented to be parallel and adjacent to one another, and cemented to a rectangle of linear sheet polarizer, resulting in a fast axis direction which changed with position across the strips. Although the assembly was divided and repeated to increase the number of strips presented to the spectrograph, only a single cycle of eight strips was used in the analysis. The assembly was mounted on a microscope slide, and protected with a front cover glass microscope slide, with a long slit fixed behind the mounting slide.

We installed the optic in a preexisting optical bench established to test the aforementioned variable retarder polarimeter, comprising an entrance aperture, a Richardson[4] transmission grating and a QSI charge-coupled device (CCD) camera. Test exposures were obtained using white light, left and right circularly polarized light produced by 3D cinema glasses, a cholesteric liquid crystal (CLC) circular polarizing filter, linearly polarized light, and a maple leaf. Fig. 2 (a) shows a section of the processed data ratioed to the average white light spectrum, while Fig. 2 (b) shows the raw data which emerged from the spectrograph. The dark gaps in the raw data correspond to the joins

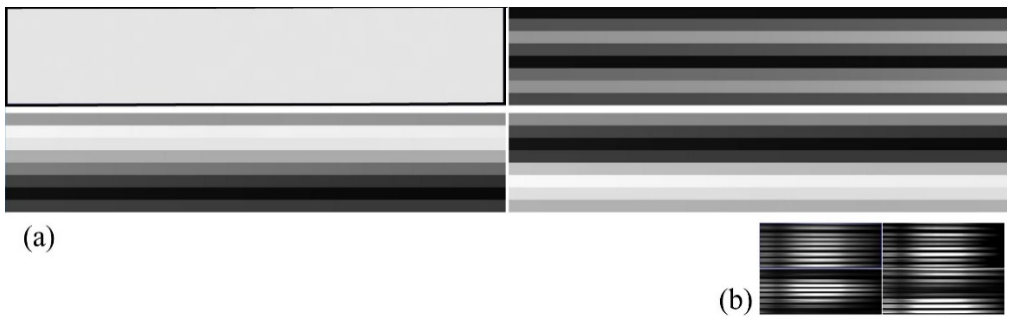

(a)

(b)

Fig. 2 (a) Main panel showing four data frames each derived using a single slit with eight quarter wave fast axis position angles, changing by approximately 22.5° between spectra after removal of gaps and normalization by continuum; top left white light, top right horizontally linearly polarized light, lower left is left-circularly polarized light and lower right is right-circularly polarized light. As expected, we see twice the spatial modulation frequency for the linearly polarized light relative to the circularly polarized light, and the expected sign change between left and right circularly polarized light. Insert (b) shows the raw data, with gaps between stripes where strips adjoin, layout as for (a). The wavelength decreases from left to right and ranges from approximately 670 nm to 590 nm shown, with a full range of about 750 nm to 450 nm.

---

between adjacent strips. As a device to measure circular polarization, note that (a) the modulation amplitude for $V$ is twice that of $Q$ or $U$ for the same degree of polarization, also reflected in the equations in the Appendix, and (b) since $V$ modulates with half the spatial frequency to the linear terms, there is, in principle, no cross-talk between the two.

Figs. 3, 4, and 5 show the results of applying the equations in Appendix A to the extracted processed spectra. Fig. 3 shows left and right circularly polarized light obtained by directing the input beam through the left and right lenses of polarization-based 3D cinema glasses. The glasses also reveal a significant amount of linear polarization, as was also found in Sparks et al. (2012). Fig. 4 shows the polarization spectrum of a cholesteric liquid crystal circularly polarizing filter, together with an inset from the original specification material. Note that the ripples present in the data are also present in the manufacturer's specifications. Finally, we passed the input beam through a maple leaf and found a significant level of circular polarization with a shape similar to previous measurements of a leaf (Sparks et al. 2009; Patty et al. 2017). The root-mean-square noise level for a single white light frame in Stokes $V$ is less than 0.2 %, also included in Fig. 5, a

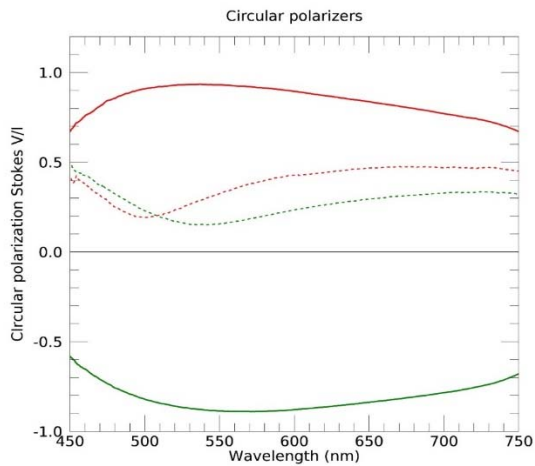

Fig. 3. Derived polarization for left and right circular polarization light obtained using white light shone through an appropriately oriented pair of 3D cinema glasses. The solid lines give the derived circular polarization and the dotted lines the linear. Green is for the left lens and red for the right. The results are similar to those shown in Sparks et al. (2012).

surprisingly high-quality performance given the inexpensive and rudimentary nature of the device. However, given the absence of polarimetric calibration, these absolute levels are quite uncertain, although the qualitative behavior is as expected. The device has satisfied its purpose of providing a proof of concept with the expected polarization sensitivities and functional behavior predicted by theory.

Returning to a more sophisticated approach, if a device were to be constructed using liquid crystal technology, an example implementation could be as follows. A common detector pixel size is 15 μm; a strip at a particular fast axis angle might cover say 4 detector pixels, or 60 μm for a 1:1 mapping between the slit and detector, well within the capabilities of current manufacturing technology (e.g. the minimum beam waist of the liquid crystal writing system of Miskiewicz & Escuti 2014 is 1$\mu$m, and for current liquid crystal technology, the "strips" are likely to flow into one another to form a truly continuous variation of fast axis direction). To span a cycle using eight angles, as here, repeated twice, i.e., two full cycles, equivalent to the vortex "charge" $m$=2, the optic would be close to 1 mm in length (0.96 mm to be precise). An $m$=4 device would be 2 mm in extent. The equivalent extent on the detector is 64 or 128 pixels respectively, in the polarimetric direction. The spatial modulation frequency and desired coverage in the polarimetric dimension (*y,* say) is a free parameter, independent of the spectroscopic configuration in the orthogonal direction (*x,* say). For example, if it is necessary to collect a very large number of photons on a single frame to provide precision polarimetry, then spreading the light more extensively in *y* may be appropriate. This would average over detector artifacts and pixel response variations. In wavelength, either a broad range to span the slowly varying spectrum of an aerosol having a preferred particle size, or a high-resolution analysis of a specific atmospheric absorption line could equally well be accommodated given the achromaticity of the polarimetric spatial modulation.

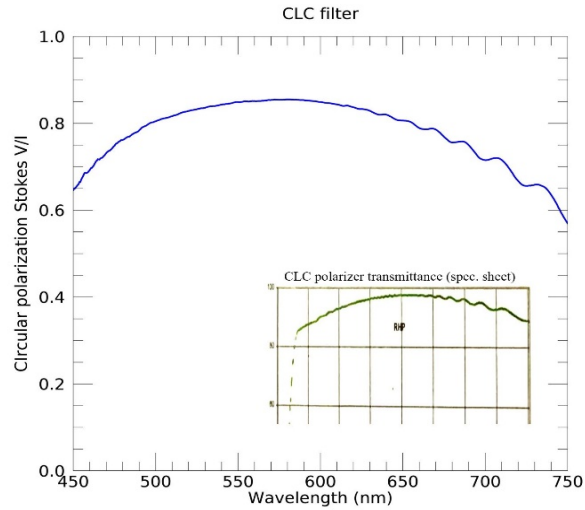

Fig. 4. Derived polarization for a cholesteric liquid crystal circularly polarizing filter. Note the ripples to the upper right, also visible in the manufacturer's specification sheet, copied in the inset.

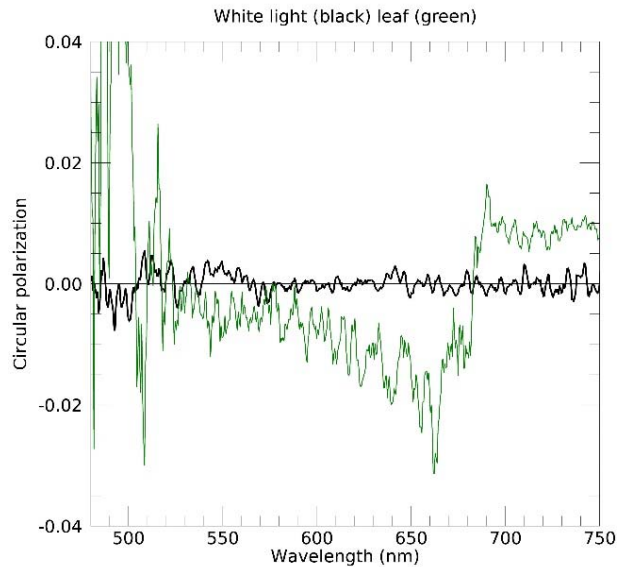

Fig. 5. Circular polarization spectrum of light transmitted through a maple leaf, showing (in green) the characteristic structure associated with the chlorophyll absorption band at about 670 nm. For comparison, the black line shows the analysis for an unpolarized white light. The root mean square of the latter is 0.19 % in Stokes *v*.

## 4. Discussion

A quarter wave plate polarimeter with a spatially variable fast axis has inherent geometric achromaticity, and circular and linear Stokes parameters are carried on spatial frequencies that differ by a factor two. Hence the configuration is robust against cross-talk between linear and circular polarization in the event of calibration errors and temperature changes. These qualities make the instrument concept an attractive one for application to remote sensing of homochirality through the circular polarization spectrum (Patty et al. 2018). The advantages relative to the spatially variable retardance polarimeter (double wedges) specifically are its achromaticity, reduced scattered light from component surfaces, and simpler sine and cosine trigonometric encoding with factor two frequency difference, as opposed to the linear and circular Stokes parameters being coefficients of more complex orthogonal trigonometric functions. The disadvantages are more complex manufacture (for the liquid crystal version), a limited wavelength range over which the wave plate is effective, compared to the far ultraviolet (FUV) to midinfrared (MIR) range of the birefringent wedges, and in the quarter wave implementation, the intensity spectrum and one of the linear Stokes parameters have a mutual dependence, which can, in certain implementations, propagate to the other Stokes parameters through the intensity normalization. Both implementations encode the full Stokes spectrum onto a single data frame and share the advantages of compactness, robustness and no moving parts.

Generalizing the concept, the rotation of essentially any polarization optic may be replaced with a static mapping onto a fixed spatial dimension, potentially eliminating the need for moving parts in a number of different polarimetric measurement approaches. Goldstein (2011) §16 discusses various methods for measuring the Stokes vector which involve rotating components, either rotating retarders or rotating analyzers, many of which are amenable to the (static) geometric approach, see Appendix B. The geometric achromaticity also means that in general, fringes are not washed out by decreasing spectral resolution, or widening of a spectral bandpass, as they are for fringe spacing which scales like $1/\lambda$, arising from a dependence on the birefringence of the polarizing component. Hence, certain aspects of instrument design are simplified and the potential to work with a broad spectral bandpass is improved.

Static encoding of the output from a quarter wave plate polarimeter could be accomplished in other ways. The polarization analyzer could be implemented using crystal optics such as a Wollaston prism, or Savart plate. Use of such components could potentially allow more effective use into the ultraviolet or infrared, while use of a beam splitter to yield a dual beam version of the polarimeter also recovers the factor of $\sqrt{2}$ in the signal-to-noise ratio calculation presented in Appendix A, as well as providing a redundant analysis and ability to remove certain systematic error terms. In particular this can help eliminate the $Q/I$ dependency of the first term in equations (A2), (A4) by reversing the sign of $Q$ (and the other Stokes parameters). Elimination of other sources of systematic uncertainty follow in similar fashion to a classical dual beam polarimeter (Patat et al. 2006; Bagnulo et al. 2009; Snik & Keller 2013), including slit illumination non-uniformity, see Appendix A.

A conical Fresnel rhomb could provide an alternative to the quarter wave retarder with spatially varying fast axis direction, yielding a similar polarization modulation on an output circle instead of a straight spatial axis when used together with a fixed position angle analyzer. Progressing around the output circumference of the rhomb, at the (wider) base of the cone, the fast axis direction remains radial and hence rotates with azimuthal position. The analyzer, meanwhile, is presumed to remain fixed in orientation, hence the net combination produces the same retarder/analyzer mix, and a continuous modulation, mapped onto a circle instead of a straight line. The modulated output may be fed to a spectrograph using optical fibers or lenses.

The twisted retarder configuration is close to that employed in so-called polarization gratings, where the incoming beam is split by the grating into left and right circularly polarized light and diffracted (Oh & Escuti 2008). We are in a regime where the spatial period of the modulation is very wide, and diffraction is not an issue. However, if the retarder were constructed with a much smaller modulation period, diffraction would require attention.

While the intent has been to propose the use of components with spatially variable geometry using liquid crystal technology or something similar, to provide the equivalent of a rotating wave plate or analyzer, we can ask whether it makes sense to consider an inexpensive implementation similar to the sheet polymer retarders used in our illustration. With machine tooling, and very inexpensive materials, it would certainly be feasible in principle to assemble a large number of devices at low cost, offering full Stokes, or just linear, polarimetry, cut from polymer sheets or manufactured in strips. Research would be required to assess the impact of light scattering from the strip edges, and the degree of difficulty to calibrate individual devices to a useful level, in conjunction with similarly inexpensive spectroscopic and detector optics in order to understand whether or not this would be a cost-effective option for polarimetry.

From a scientific perspective, the advantages of enabling full Stokes polarimetry on a single data frame without sequential data acquisition are substantial. Transient phenomena whose polarization might be measured in this way include: Solar and stellar flares, M dwarf flares and light echoes (Sparks et al. 2018), exoplanet transit polarization, aurorae, lightning, fast explosions and gravitational lens events to name a few. Similarly, if the data are to be obtained from a moving platform, or the target is moving relative to the instrumentation, then the polarimetric data will represent the average over the duration of an exposure, and will not be compromised by changes occurring on that or other timescales. For the purposes of life detection through remote sensing of chirality (Sparks et al. 2009; Sparks et al. 2015; Patty et al. 2018), the sensitivity of such a device to circularly polarized light coupled to the complete lack, in principle, of cross-talk between linear and circular polarizations make it an efficient, powerful and hence attractive option for this application.

## 5. Conclusion

By mapping the time dimension or rotation angle of a rotating wave plate polarimeter onto a spatial dimension by using e.g. liquid crystal manufacturing methods, or simply by assembling such a device macroscopically, it is possible to reconfigure a classical approach to polarimetry and spectropolarimetry to acquire data on a single two-dimensional data frame. The advantages are compact design, robustness, increased fidelity for time-variable or moving targets, and no moving parts. This enables polarimetric measurement of extremely rapid phenomena, and eliminates degradation of data due to changes through a sequential acquisition. A quarter wave plate configuration is a very effective tool for measuring the circular polarization of a target, since it uses two different carrier frequencies for circular and linear polarization, thereby eliminating in principle, any cross-talk between linear and circular polarizations, and has maximal sensitivity to the circular polarization. Such a device has application in remote life sensing, as the circular polarization spectrum can be used to remotely sense the presence of chiral, biological material.

**Acknowledgement**

We thank the reviewer, Frans Snik, for a thorough and constructive review which improved the paper significantly.

**Appendix A: Mueller matrix analysis for quarter wave polarimeter**

*Solution for the Stokes parameters*

The use of a quarter wave polarimeter is widespread, with an extensive literature. The optimal analytic solutions for polarimetric demodulation in general are presented by del Torro Iniesta & Collados (2000). Here, for convenience and consistency with our previous approach (Sparks et al. 2012), we summarize the relevant equations using a Mueller matrix formalism and linear system solution. Goldstein (2011), for example, presents an equivalent Fourier series approach and solution for the rotating quarter wave plate polarimeter. Let an incoming light beam be described by its Stokes vector $S = (I, Q, U, V) \equiv I(1, q, u, v)$ where $I$ is the total intensity, $Q$, $U$, $V$ are the unnormalized Stokes parameters, and $q, u, v$ are the normalized Stokes parameters. $I$ is the total intensity; $Q$, $U$ describe the linear polarization and $V$ the circular. After transmission through a wave plate with fast axis at angle $\theta$, and a polarization analyzer, the observed intensity is distributed (simultaneously) across $n$ positions corresponding to a set of fast-axis angles $\{\theta_i\}$. At each fast axis angle, the effective input Stokes vector is $S' = (I, Q, U, V)/n \equiv (I/n)(1, q, u, v)$.

Following Goldstein (2011), the Mueller matrix for a retarder (wave plate) with retardance $\phi$ and fast axis at angle $\theta$ to the $x$ axis is:

$$M_W(\phi, \theta) = \begin{pmatrix} 1 & 0 & 0 & 0 \\ 0 & \cos^2 2\theta + \cos\phi \sin^2 2\theta & (1 - \cos\phi) \sin 2\theta \cos 2\theta & -\sin\phi \sin 2\theta \\ 0 & (1 - \cos\phi) \sin 2\theta \cos 2\theta & \sin^2 2\theta + \cos\phi \cos^2 2\theta & \sin\phi \cos 2\theta \\ 0 & \sin\phi \sin 2\theta & -\sin\phi \cos 2\theta & \cos\phi \end{pmatrix}.$$

The Mueller matrix for an imperfect polarizer is:

$$M_p(t, \alpha) = \frac{t^2}{2} \begin{pmatrix} 1 & \cos 2\alpha & 0 & 0 \\ \cos 2\alpha & 1 & 0 & 0 \\ 0 & 0 & \sin 2\alpha & 0 \\ 0 & 0 & 0 & \sin 2\alpha \end{pmatrix},$$

where the transmittance of the electric vector in the $x$ and $y$ directions is $t_x$ and $t_y$ respectively, $t^2 = t_x^2 + t_y^2$, and $\alpha = \tan^{-1}(t_y/t_x)$. For an ideal or perfect horizontal polarizer, $\alpha = 0$ and $t = 1$. Hence the output intensity profile for the incoming beam after encountering these two optical elements at each angle $\theta_i$ is

$$I_{obs}(\phi, \theta_i) = \frac{t^2}{2n} \Big( I + \frac{1}{2} \cos 2\alpha (1 + \cos\phi) Q \\ + \frac{1}{2} \cos 2\alpha (1 - \cos\phi) Q \cos 4\theta_i \\ + \frac{1}{2} \cos 2\alpha (1 - \cos\phi) U \sin 4\theta_i \\ - \cos 2\alpha \sin\phi\, V \sin 2\theta_i \Big). \tag{A1}$$

For an ideal linear polarizer, $t = 1$ and $\alpha = 0$. For an ideal quarter wave plate, $\phi = \pi/2$ and for an ideal half wave plate $\phi = \pi$.

Hence, the intensity modulation for an ideal quarter wave plate plus ideal (single element) polarizer is (e.g., Goldstein 2011):
$$I_{obs}(\phi, \theta_i) = \frac{1}{2n}\left(I + \frac{1}{2}Q + \frac{1}{2}Q\cos 4\theta_i + \frac{1}{2}U\sin 4\theta_i - V\sin 2\theta_i\right). \tag{A2}$$

Similarly, the ideal half wave plate plus ideal (single element) polarizer is:
$$I_{obs}(\phi, \theta_i) = \frac{1}{2n}(I + Q\cos 4\theta_i + U\sin 4\theta_i). \tag{A3}$$

Treating equation (A2) as a linear system of the form
$$I_{obs}(\theta_i) = a + b\cos 4\theta_i + c\sin 4\theta_i + d\sin 2\theta_i \tag{A4}$$

and solving for $a, b, c, d$, the Stokes parameters are given by:
$$\begin{aligned} I &= 2n(a-b) \\ Q &= 4nb; \quad q = Q/I = 2b/(a-b) \\ U &= 4nc; \quad u = U/I = 2c/(a-b) \\ V &= -2nd; \quad v = V/I = -d/(a-b). \end{aligned} \tag{A5}$$

To solve the system, we use a standard least squares analysis. Denoting $I_{obs}(\phi, \theta_i)$ as $I_i$, then $\chi^2 = \sum_i \frac{1}{\sigma_i^2}(I_i - a + b\cos 4\theta_i + c\sin 4\theta_i + d\sin 2\theta_i)$ where $\sigma_i$ is the uncertainty on measurement $I_i$. Minimizing $\chi^2$ for each of $a, b, c, d$:

$$\frac{1}{\sigma_i^2}(\sum I_i, \quad \sum I_i \cos 4\theta_i, \quad \sum I_i \sin 4\theta_i, \quad \sum I_i \sin 2\theta_i) = \boldsymbol{B} \cdot \begin{pmatrix} a \\ b \\ c \\ d \end{pmatrix},$$

$$\boldsymbol{B} = \frac{1}{\sigma_i^2}\begin{pmatrix} n & \sum \cos 4\theta_i & \sum \sin 4\theta_i & \sum \sin 2\theta_i \\ \sum \cos 4\theta_i & \sum \cos^2 4\theta_i & \sum \cos 4\theta_i \sin 4\theta_i & \sum \cos 4\theta_i \sin 2\theta_i \\ \sum \sin 4\theta_i & \sum \cos 4\theta_i \sin 4\theta_i & \sum \sin^2 4\theta_i & \sum \sin 4\theta_i \sin 2\theta_i \\ \sum \sin 2\theta_i & \sum \cos 4\theta_i \sin 2\theta_i & \sum \sin 4\theta_i \sin 2\theta_i & \sum \sin^2 2\theta_i \end{pmatrix},$$

(where for clarity we have assumed all the error terms $\sigma_i^2$ are the same). Hence (see also Goldstein 2011, equations 14.70-14.86):

$$(a, \quad b, \quad c, \quad d) = \boldsymbol{B}^{-1} \cdot \begin{pmatrix} \sum I_i \\ \sum I_i \cos 4\theta_i \\ \sum I_i \sin 4\theta_i \\ \sum I_i \sin 2\theta_i \end{pmatrix}.$$

*Error analysis*

The variances in $a, b, c, d$ are $\sigma_a, \sigma_b, \sigma_c, \sigma_d$ given by the diagonal terms of the inverse curvature matrix $\boldsymbol{B}^{-1}$. The functions $\sin 4\theta_i, \cos 4\theta_i \sin 2\theta_i$ are mathematically orthogonal, hence the cross-terms integrate to zero over an integer number of periods in $\theta$. For an analytic approximation to the uncertainties, we neglect the off-diagonal terms and approximate the quadratic sums in $\boldsymbol{B}$ from the equivalent integral over the range $\Delta\theta$: $\sum \cos^2 4\theta_i (\Delta\theta/n) \approx \int_0^{\Delta\theta} \cos^2 4\theta_i \, d\theta = \Delta\theta/2$, similarly the other terms. Hence $\sum \cos^2 4\theta_i \approx n/2$ where $n$ is the number of measurements. Hence

$$\boldsymbol{B^{-1}} \approx \sigma_i^2 \begin{pmatrix} 1/n & 0 & 0 & 0 \\ 0 & 1/\sum \cos^2 4\theta_i & 0 & 0 \\ 0 & 0 & 1/\sum \sin^2 4\theta_i & 0 \\ 0 & 0 & 0 & 1/\sum \sin^2 2\theta_i \end{pmatrix} \approx \frac{\sigma_i^2}{n} \begin{pmatrix} 1 & 0 & 0 & 0 \\ 0 & 2 & 0 & 0 \\ 0 & 0 & 2 & 0 \\ 0 & 0 & 0 & 2 \end{pmatrix}.$$

If the total number of detected photons in the observations is $N_{tot} = \sum I_i$ (which for an ideal single polarizer in the system means the input Stokes $I \approx 2\,N_{tot}$), then $\sigma_i \approx \sqrt{N_{tot}/n}$ since typically polarization degree is small and hence the $I_i$ values are approximately the same. Therefore,

$$\boldsymbol{B^{-1}} \approx \frac{N_{tot}}{n^2} \begin{pmatrix} 1 & 0 & 0 & 0 \\ 0 & 2 & 0 & 0 \\ 0 & 0 & 2 & 0 \\ 0 & 0 & 0 & 2 \end{pmatrix},$$

i.e., $\sigma_a^2 \approx N_{tot}/n^2$, $\sigma_b^2 \approx 2\,N_{tot}/n^2 \approx \sigma_c^2 \approx \sigma_d^2$. Inserting into the solution equations, (A5), $I = 2n(a-b)$ hence $\sigma(I) \approx 2n\sqrt{\sigma_a^2 + \sigma_b^2} \approx 2\sqrt{3}\sqrt{N_{tot}}$. For low polarization degree $\sigma(Q) \approx 4\sqrt{2}\sqrt{N_{tot}}$, $\sigma(q) \approx \sigma(Q)/I \approx 2\sqrt{2}/\sqrt{N_{tot}}$.

$$\begin{aligned} \sigma(q) &\approx 2\sqrt{2}/\sqrt{N_{tot}} \\ \sigma(u) &\approx 2\sqrt{2}/\sqrt{N_{tot}} \\ \sigma(v) &\approx \sqrt{2}/\sqrt{N_{tot}}. \end{aligned} \qquad (A6)$$

*Flat fielding and slit illumination errors*

The error on a single pixel measurement is $\sigma_i$ which for photon noise is $\sqrt{I_i}$, leading to the equations (A6). If in addition there is an independent error on the flat field response of the pixel, $\sigma_f$ say, then $\sigma_i = \sqrt{(I_i + I_i^2\,\sigma_f^2)}$, hence flat field errors will dominate over photon errors when $\sigma_f > 1/\sqrt{I_i}$. Now $I_i \approx N/n$ where n is the number of pixels in the polarimetric dimension (and any wavelength binning) and $N$ is the number of collected photons, hence this condition becomes $\sigma_f > \sqrt{(n/N)}$. In the limit for $n=1$ (no spreading of the beam), this can be very stringent. For example, if we require a polarization accuracy of ~$10^{-4}$ polarization degree, we need ~$10^8$ photons, and the flat field needs to be accurate to 1 part in $10^4$ (in the absence of mitigating measures from a dual beam or related strategies). On the other hand, if the beam is spread over say 100 pixels, this is relaxed to 1 part in $10^3$, which is within the capabilities of a straightforward calibration system. This illustrates the improved robustness against flat-field errors obtained by spreading the beam across multiple pixels.

If there is non-uniform illumination of the entrance slit, $S(\theta)$, which influences the output intensity multiplicatively, the Fourier terms of $S(\theta)$ could mimic source polarization. In the single beam implementation, care will be required in design to minimize non-uniformity of slit illumination, and in calibration to account for any such residual variation. By working with a dual beam version, however, reversing the signs of the Stokes parameters in (A2) for the second beam, if the ordinary and extraordinary beams are given by $I_o$ and $I_e$, by working with normalized flux differences $f=(I_o-I_e)/(I_o+I_e)$, following Patat et al. (2006), or flux ratios following Miller et al. (1988), the slit function $S(\theta)$ cancels since it is common to both beams and the slit non-uniformity dependence is removed. Additionally, in the dual beam configuration the modulation equation (A2) becomes an equation in only the normalized Stokes parameters, *q,u,v*, removing the "*I*" dependency, and taking both beams together means that no significant light is lost, improving *S/N* by $\approx\sqrt{2}$.

## Tolerance analysis

### Retardance not exactly π/2

If the quarter wave retarder instead provides a retardance $\phi=\pi/2-\varepsilon$, then for small $\varepsilon$ for an ideal polarizer ($t=1$) equation (A1) becomes:

$$I_{obs}(\phi,\theta_i) = \frac{1}{2n}\left(I + \frac{1}{2}(1+\varepsilon)Q \right.$$
$$+ \frac{1}{2}(1-\varepsilon) Q \cos 4\theta_i \quad\quad (A7)$$
$$+ \frac{1}{2}(1-\varepsilon) U \sin 4\theta_i$$
$$\left. - \left(1 - \frac{\varepsilon^2}{2}\right) V \sin 2\theta_i \right).$$

It may be seen from these expressions that only a small correction to the derived values of $Q$, $U$ and $V$ is required if $\varepsilon$ is small. For $Q$, $U$ the correction factor is $\approx(1+\varepsilon)$ while for $V$ it is a second order correction $\approx(1 + \varepsilon^2/2)$. Alternatively, if $\varepsilon$ is not necessarily small, the sensitivities and variances between the Stokes parameters can be manipulated by changing the retardance. No cross-talk is introduced between the Stokes parameters by a slightly imperfect quarter wave retarder, nor by using a general retardance as in (A1).

### Angles not exactly as expected

For the case where the retarder angles are offset from the expected angles by a small amount $\delta$, the true angle $\theta_i = \theta_i' + \delta$. Substituting into equation (A1), and collecting $\cos 4\theta_i$, $\sin 4\theta_i$ terms, to first order in $\delta$:

$$I_{obs}(\phi,\theta_i) = \frac{t^2}{2n}\left(I + \frac{1}{2}\cos 2\alpha(1 + \cos\phi)Q \right.$$
$$+ \frac{1}{2}\cos 2\alpha(1 - \cos\phi)(Q + 4U\delta)\cos 4\theta'_i \quad\quad (A8)$$
$$+ \frac{1}{2}\cos 2\alpha(1 - \cos\phi)(U - 4Q\delta)\sin 4\theta'_i$$
$$- \cos 2\alpha \sin\phi\, V \sin 2\theta'_i$$
$$\left. - 2\cos 2\alpha \sin\phi\, V \cos 2\theta'_i\, \delta \right).$$

In this case, by incorrectly calibrating the retarder angles, we do see crosstalk between the linear Stokes parameters $Q$ and $U$, with ($Q+U\,\delta$) replacing the correct $Q$ in the coefficient of $\cos 4\theta_i$ with similar cross-talk for $U$. However, the coefficient of $\sin 2\theta_i$ is unchanged, and hence the value of $V$ is robust against such an error. The presence of a $\cos 2\theta_i$ term would be a warning that an angle miscalibration may be present.

### Inefficient polarizer

Equation (A1) encapsulates the impact of inefficient polarizers, with the parameters *t*, the polarizer transmission, and $\alpha$ giving the relative *x* and *y* transmissions. The form of the equation is the same (A4), with the coefficient of the solution $Q=4nb/(t^2\cos 2\alpha)$, $U=4nc/(t^2\cos 2\alpha)$, and $V = -2nd/(t^2\cos 2\alpha)$ and $I=2n(a-b)/t^2$, which gives $q=2b/((a-b)\cos 2\alpha)$, $u=2c/((a-b)\cos 2\alpha)$, $v=-d/((a-b)\cos 2\alpha)$, i.e., as before corrected for the polarizer inefficiency given by $1/\cos 2\alpha$, with the total intensity to be corrected by the transmission factor $1/t^2$. Once again, this introduces no cross-talk, only scaling factors related to the efficiency of the polarizer component.

# Appendix B: Example methods for Stokes polarimetry

We present a sample selection of rotating component polarimeters which could be reconfigured using geometric manipulation of retarder and analyzer angles to acquire spectroscopic data on a single two-dimensional data frame, to illustrate the generality of the approach, see Table 1. The list is not intended to be complete, but illustrative. Goldstein (2011) § 16 contains multiple rotating component configurations for Stokes polarimetry, see his figure 16.2 for example.

TABLE 1: Examples of rotating component polarimeters which could encode the polarization along a static polarimetric dimension.

| Description | Modulation (single beam version) | Example equivalent | Comments |
|---|---|---|---|
| Rotating polarizer/analyzer only | $I_{obs}(\theta_i) = \frac{1}{2n}(I + Q \cos 2\theta_i + U \sin 2\theta_i)$ | HST/ACS linear polarizers at 60° angles | G16.5[5] simplest rotating analyzer configuration |
| Rotating half wave plate plus fixed analyzer | $I_{obs}(\theta_i) = \frac{1}{2n}(I + Q \cos 4\theta_i + U \sin 4\theta_i)$ | Typical mode for astronomical dual beam polarimeter | |
| Rotating quarter wave plate plus fixed analyzer | $I_{obs}(\theta_i) = \frac{1}{2n}(I + \frac{1}{2}Q + \frac{1}{2}Q \cos 4\theta_i + \frac{1}{2}U \sin 4\theta_i - V \sin 2\theta_i)$ | Quarter wave polarimeter discussed in the text | Circular and linear use carrier frequencies different by factor 2. G14.64 |
| Rotating quarter wave retarder plus rotating analyzer | $I_{obs}(\theta_i, \phi_i) = \frac{1}{2n}(I + \frac{1}{2}Q \cos 2\theta_i + \frac{1}{2}U \sin 2\theta_i \cos \phi_i + V \sin 2\theta_i \sin \phi_i$ | $\theta$ is quarter wave retarder fast axis angle and $\phi$ is analyzer axis angle | Stokes approach to Mueller matrix measurement. G14.5 |

---

[5] GXX.YY refers to the equation number of Goldstein (2011).